\documentclass[%
 reprint,
 amsmath,amssymb,
 aps,
nofootinbib]{revtex4-1}
\usepackage{booktabs}
\usepackage{parskip}
\usepackage{graphicx}
\usepackage{dcolumn}
\usepackage{bm}
\usepackage[symbol]{footmisc}

\usepackage{amsmath}
\usepackage{xcolor}
\usepackage{multirow}
\usepackage{chemformula}

\begin{document}

\preprint{APS/123-QED}

\title{Reduced-Order Modelling of Defect Transport using Surrogate Kinetics: Application to U$_x$Pu$_ {1-x}$N}

\author{Peter Hatton}
\affiliation{Amentum Clean Energy, 305 Bridgewater Place, Birchwood Park, Warrington, UK, WA3 6XF}

\date{\today}

\begin{abstract}
Defect transport in chemically disordered materials is a difficult phenomenon to model since migration energetics depend strongly on the local chemical environment, producing a distribution of transition barriers that cannot be exhaustively enumerated. Here, we develop a reduced-order approach to modelling defect diffusion in compositionally complex materials using environment-dependent surrogate kinetics. Migration energetics for actinide vacancies, nitride vacancies, and actinide–nitride divacancies in U$_x$Pu$_{1-x}$N are generated using the Hop-Decorate workflow across thousands of chemical configurations. These data are distilled into compact surrogate functions that predict migration barriers and energy differences from simple descriptors based on local coordination counts. The surrogate models reproduce the atomistic dataset with low error and enable efficient evaluation of migration rates during lattice kinetic Monte Carlo simulations. Long-time diffusivities computed across temperature and composition reveal strongly non-linear behaviour arising from two dominant mechanisms: species-controlled migration on the actinide sublattice and environment-dependent trapping on the nitride sublattice. Although demonstrated for U$_x$Pu$_{1-x}$N the framework provides a general and computationally efficient approach for modelling defect transport in chemically disordered materials and for integrating atomistic kinetics into higher-scale simulations.
\end{abstract}

\maketitle

\section{Introduction}

The generation, migration and interaction of point defects are central processes in materials science, controlling many of the mechanisms of microstructural evolution which lead to changes in macroscopic material properties and potentially materials failure under extreme conditions. Indeed, point defect transport directly influences phenomena such as swelling, creep, radiation-induced segregation, and changes in thermal or mechanical performance \cite{Nastar2012,Nichols1978,Rajasekaran2016,Xiao1995,Faulkner1980}. Therefore, understanding and predicting defect diffusivity is a fundamental requirement across fields ranging from metallurgy to semiconductor device engineering, and is especially critical in nuclear materials where irradiation continually produces and drives the migration of defects.

Numerous computational strategies exist for quantifying defect transport in crystalline systems with high symmetry and chemically uniform lattices, where defect migration pathways can be enumerated and represented as a finite transition network. Within this framework, transport can be modelled rigorously as a continuous-time Markov chain (CTMC) in which each node corresponds to a defect state and each edge represents a possible migration event with an associated rate \cite{Voter2007,LeBris2012}. In such idealised cases, defect diffusivities can be obtained analytically by solving for the long-time behaviour of the CTMC or by evaluating Green–Kubo relations for the diffusion tensor \cite{Novotny1995,van2018,Swinburne2020}. Computational tools and methods for this class of problem are well developed and have been successfully applied to a wide variety of ordered crystalline systems \cite{Swinburne2020,Ebmeyer2024,Becquart2009,Perez2016}. Beyond analytical treatments, stochastic atomistic methods provide a complementary means of sampling defect dynamics. Molecular dynamics (MD), for example, enables direct observation of defect motion at finite temperatures, yielding effective diffusivities or jump frequencies when sampling is sufficient \cite{Korfer2023,Vaari2015}. First-principles approaches, such as density functional theory (DFT), allow for accurate calculation of migration barriers and defect energetics, which can then parameterize kinetic Monte Carlo (KMC) simulations to extend accessible timescales \cite{Schneider2025}. These methods are powerful in ordered systems, but they depend fundamentally on either full enumeration of symmetry-distinct environments or sufficient dynamical sampling, both of which become intractable in the presence of chemical complexity.

The breakdown of translational and rotational symmetry in chemically complex materials, or those that develop complexity through life, lead to the generation of distinct local chemical environments through which a defect may migrate \cite{Talapatra2025}. Each environment can be associated with a unique migration barrier, resulting in a distribution of kinetics rather than a small, closed set of rates \cite{Huang2025,Xu2022,Zhao2022,Xing2026}. Full sampling of these environments via DFT or MD is computationally prohibitive, and conventional CTMC models cannot accommodate the effectively infinite variety of transition rates. Furthermore, migration itself can perturb the local chemical ordering, as in the case of vacancy diffusion where atomic displacements shuffle chemical species leading to an evolving chemical environment \cite{Hatton2023}, further complicating analytical treatments.

A material which exemplifies these challenges is that of the proposed advanced nuclear fuel, uranium mono-nitride (UN). UN has long been considered a promising candidate for advanced nuclear fuel because of its high thermal conductivity, high fissile atom density, and favourable thermomechanical properties compared to oxide fuels \cite{Ekberg2018,Uno2020,Abdulhameed2025}. UN may also exist or evolve as a chemically mixed nitride since, on the one hand, fission events and transmutation reactions during irradiation introduce actinide products such as Pu into the lattice, resulting in U$_x$Pu$_{1-x}$N solid solutions. On the other hand, mixed actinide nitrides can also be deliberately manufactured as candidate fuel forms \cite{Syarifah2015,Arai1993}. In either case, compositional variation alters the chemical environment experienced by migrating defects, thereby modifying defect energetics. Since defect transport underpins fuel swelling, fission gas release, and degradation of thermal conductivity, capturing these effects is critical for predictive fuel performance modelling \cite{Zullo2023,Peng2021}. At the same time, direct atomistic simulation of U–Pu nitrides across the vast configurational space of local environments is infeasible, and higher-level fuel performance codes require reduced-order inputs that embed atomic scale mechanisms into tractable functional forms, as has been done for static properties across the UN-PuN composition space \cite{Galvin2024}.

To address this, the manuscript proposes a workflow which allows for atomic scale mechanisms which control defect transport in chemically complex materials to be embedded in simple Arrhenius-style equations for diffusivity, as a function of both temperature and chemical composition. In Section \ref{sec:1}, defect migration data is generated for a range of representative defects in a few global U$_x$Pu$_{1-x}$N compositions and in Section \ref{sec:2} surrogate models in the form of linear functions are fitted to this data encoding the energy barrier and energy change from defect migrations. In Section \ref{sec:3}, these surrogate models are embedded in a KMC simulator to find the relationships between long-range diffusivity, temperature and global composition. Additionally, in Section \ref{sec:4}, the surrogate models are leveraged to describe the kinetic-thermodynamic space that defects experience during diffusion to assist in elucidating the state-to-state migration mechanisms of each defect.

\section{Methodology}

Defect migration data for U$_x$Pu$_{1-x}$N systems are generated using the \textit{Hop-Decorate} (HopDec) workflow \cite{Hatton2025,HD}, which is built upon the Large-scale Atomic/Molecular Massively Parallel Simulator (LAMMPS) \cite{Thompson2022} and the Atomic Simulation Environment (ASE) \cite{Hjorth2017}. Interatomic energies and forces for U$_x$Pu$_{1-x}$N are obtained through a many-bodied Finnis–Sinclair EAM potential combined with a pairwise Buckingham interatomic potential which has been specifically validated against defect migration barriers from DFT for U/N neutral vacancy and interstitial defects in UN and to static properties, such as defect formation energies, heat capacity, and thermal expansion, in mixed U$_x$Pu$_{1-x}$N systems \cite{Kocevski2022, Galvin2024}.

For each defect type, prototypical 1NN (first nearest neighbour) migration pathways are identified in the reference UN structures. These pathways are then ``re-decorated'' 1000 times at global compositions of U$_{0.5}$Pu$_{0.5}$N, U$_{0.1}$Pu$_{0.9}$N, and U$_{0.9}$Pu$_{0.1}$N. Each decorated configuration gave differing local chemical environments around the migrating defect and therefore, varying energy barriers, as illustrated in \cite{Hatton2025}. The migration energetics are evaluated using the climbing-image Nudged Elastic Band (CI-NEB) method \cite{Henkelman2000} for each decoration.

This procedure produces a database of approximately 3000 migration barriers ($E_b$) and energy differences ($\Delta E$), sampled across a wide range of local chemical environments for each migrating defect. This database is then used to construct surrogate models that predict migration energetics as a function of the local chemical environment. Here, the local environment is defined by the counts of U atoms within NN shells at specified radii around the defect in its initial, final, and saddle point configurations. The surrogate models are formulated as linear functions of these descriptors:

\begin{equation}\label{eqn-dE}
    \Delta E = \sum_k^4 (a_k \cdot \Delta\text{N}_U^{r^M_k})
\end{equation}
\begin{equation}\label{eqn-bar}
    E_b = \sum_k^4 (b_k \cdot \Delta\text{N}_U^{r^M_k} + c_k \cdot \text{N}_U^{r^S_k}) + \beta
\end{equation}

where $a_k$, $b_k$, and $c_k$ are fitted parameters, $\Delta N_\text{U}^{r^M_k}$ represents the change in uranium counts within the $k^\text{th}$ minima shell, and $N_\text{U}^{r^S_k}$ represents U counts within the $k^\text{th}$ saddle-point shell. The set of radii $\{r^M_k, r^S_k\}$ correspond to successive NN shells; their values are listed in Table~\ref{tab-rvalues} for defects on the A (U/Pu) and N sublattice. Finally, $\beta$ is a fitted correction term related to the energy barrier of the species in pure (U/Pu)N.

\begin{table}[htbp]
    \centering
    \begin{tabular}{|c|c|c|}
        \hline
        \textbf{Parameter} & \textbf{A Sublattice} & \textbf{N Sublattice} \\ \hline
         $r^M_1$ & 4.5 & 3.0 \\
         $r^M_2$ & 5.6 & 4.5 \\
         $r^M_3$ & 6.3 & 5.6\\
         $r^M_4$ & 7.4 & 6.3 \\ \hline
         $r^S_1$ & 3.5 & 2.4 \\ 
         $r^S_2$ & 4.2 & 3.5 \\
         $r^S_3$ & 4.9 & 4.2 \\
         $r^S_4$ & 5.5 & 4.9 \\
         \hline
    \end{tabular}
    \caption{Radii of nearest neighbours shells on each sublattice for the minima (M) and saddle point (S) configurations.}
    \label{tab-rvalues}
\end{table}

The fitted surrogate models thus provide computationally inexpensive predictions of $E_b$ and $\Delta E$ for arbitrary local environments, enabling rapid evaluation during KMC simulations.

The surrogate models are integrated into a custom lattice-based kinetic Monte Carlo (laKMC) algorithm designed to simulate long-timescale defect migration. At each KMC step, possible migration events out of the current state are assigned rates according to an Arrhenius relation,

\begin{equation}
    \Gamma_i = \nu \exp\left(-\frac{E_{b,i}}{k_\text{B} T}\right),
\end{equation}

where $\nu$ is the attempt frequency (taken as $10^{13}$~s$^{-1}$ for each hop), $E_{b,i}$ is the migration barrier predicted by the surrogate model for event $i$, $k_\text{B}$ is Boltzmann’s constant, and $T$ is the temperature.  

The next event is selected using a roulette algorithm, with time advancing according to $\Delta t = -ln(u)/R$ where $u \in (0,1]$ is a random number and $R$ is the sum of all event rates. Each simulation is run for 100,000 KMC steps. Defect mean-squared displacements (MSDs) are extracted from the trajectories, and diffusivities are calculated via

\begin{equation}
    D = \frac{1}{6} \frac{d}{dt}\langle r^2(t)\rangle,
\end{equation}

Simulations are performed on supercells of size 10$\times$10$\times$10 nm$^3$ containing 64,000 atoms with periodic boundaries and the cell dimensions held fixed. For each composition, five independent random distributions of U and Pu are generated, and ten trajectories are computed per distribution, yielding a total of 50 trajectories per temperature and composition for statistical averaging. In the limiting cases of pure UN and PuN, the migration networks reduce to simple CTMCs with a single unique defect state and transition. In these cases, diffusivity can be obtained analytically \cite{Swinburne2020}.

\section{Results \& Discussion}

\subsection{Data Generation \& Analysis}\label{sec:1}

Three defect species are considered in this study: actinide vacancy (V$_\text{A}$), nitride vacancy (V$_\text{N}$), and an actinide-nitride vacancy (V$_\text{A-N}$) pair; a so-called Schottky defect. These defects are selected as representative intrinsic defects in stoichiometric UN \cite{Li2023}. A detailed characterisation of their migration energetics across U$_x$Pu$_{1-x}$N compositions is therefore essential for higher length-scale models of fuel evolution.

\begin{figure}
    \centering
    \includegraphics[width=\linewidth]{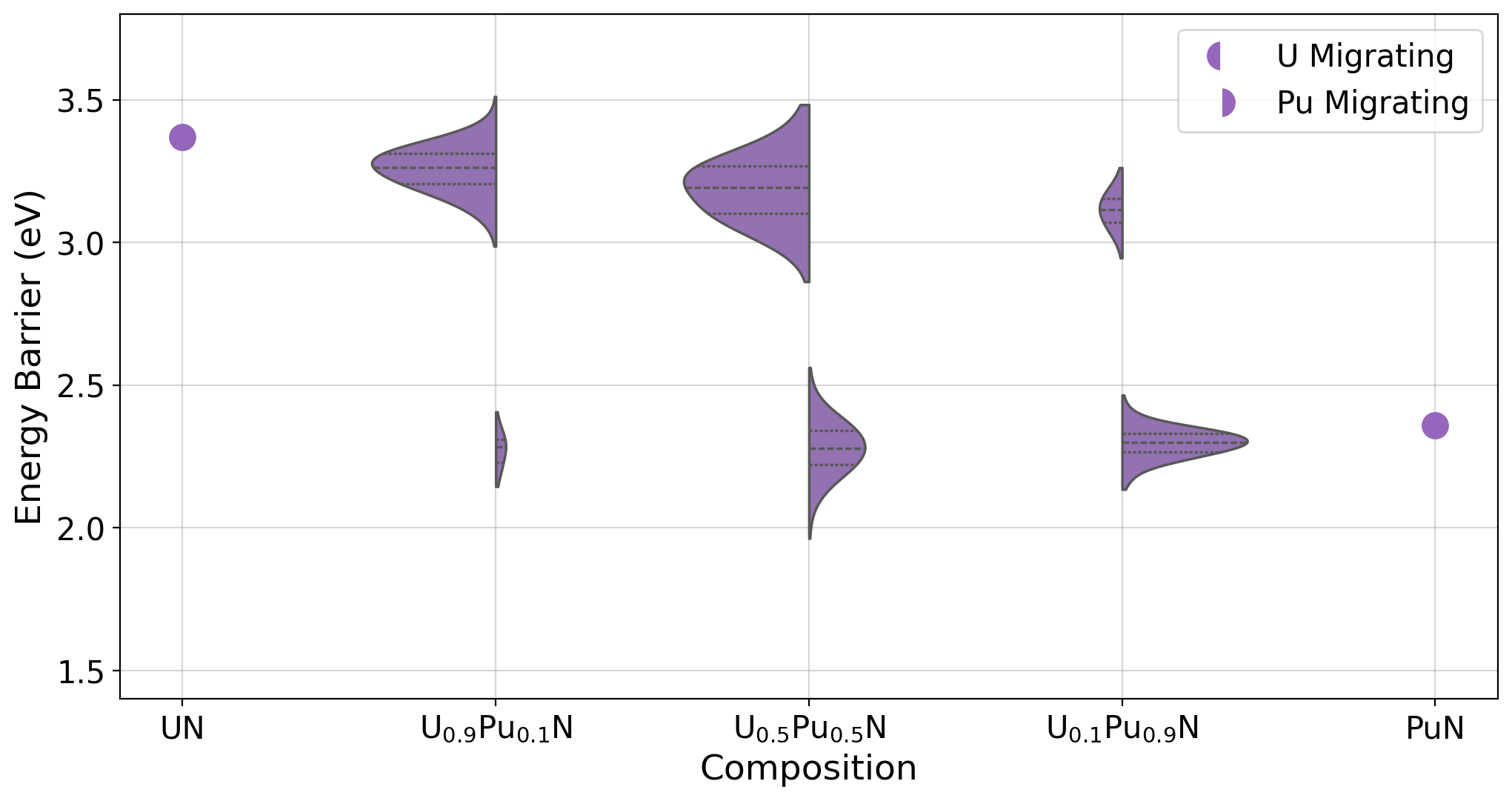}
    \caption{Energy barrier distribution for actinide vacancy migration in structures with varying global concentrations of U/Pu derived from HopDec.}
    \label{fig:U-defect-data}
\end{figure}

\begin{figure}
    \centering
    \includegraphics[width=\linewidth]{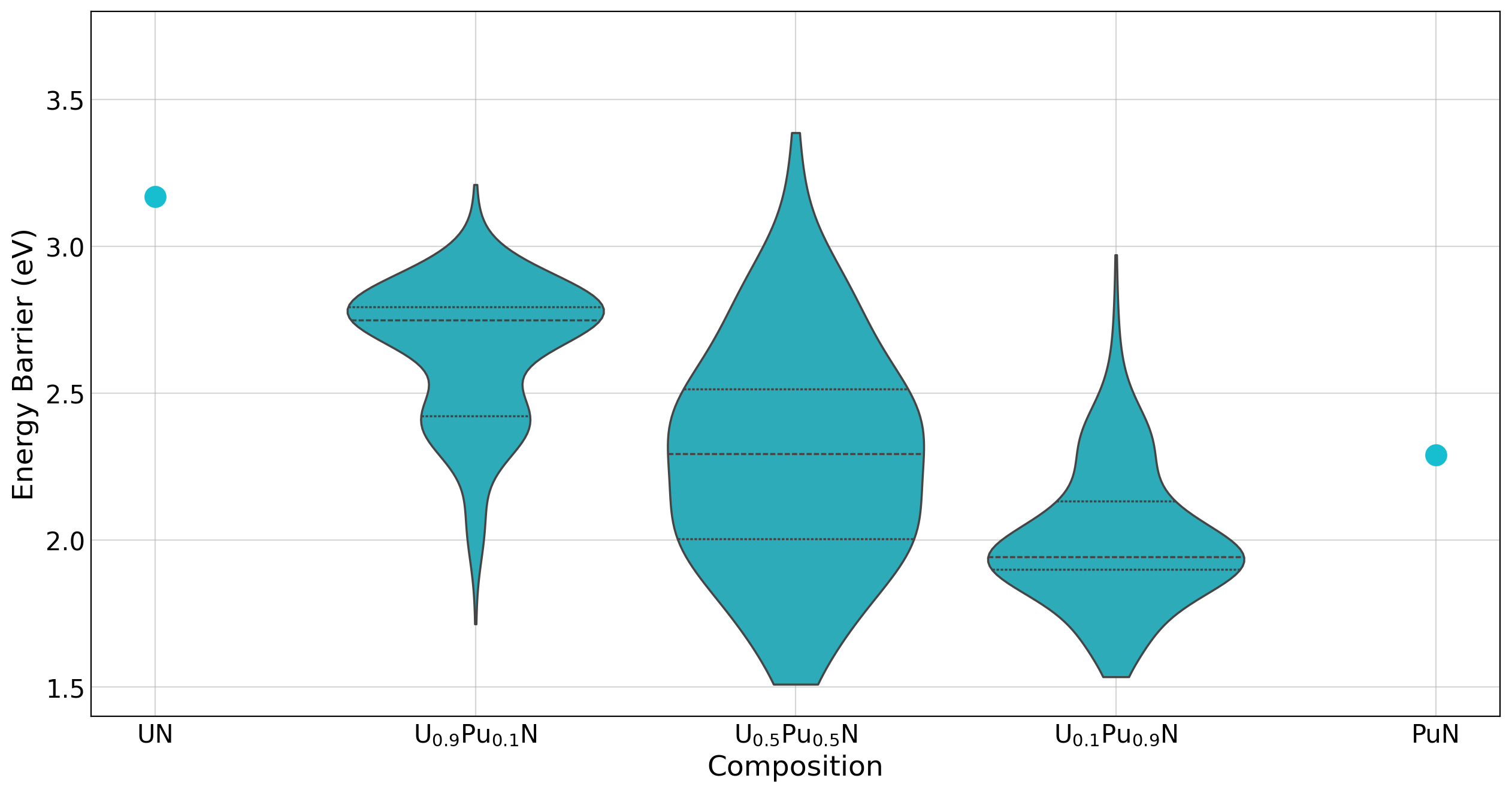}
    \caption{Energy barrier distribution for nitride vacancy migration in structures with varying global concentrations of U/Pu derived from HopDec.}
    \label{fig:N-defect-data}
\end{figure}

\begin{figure}
    \centering
    \includegraphics[width=\linewidth]{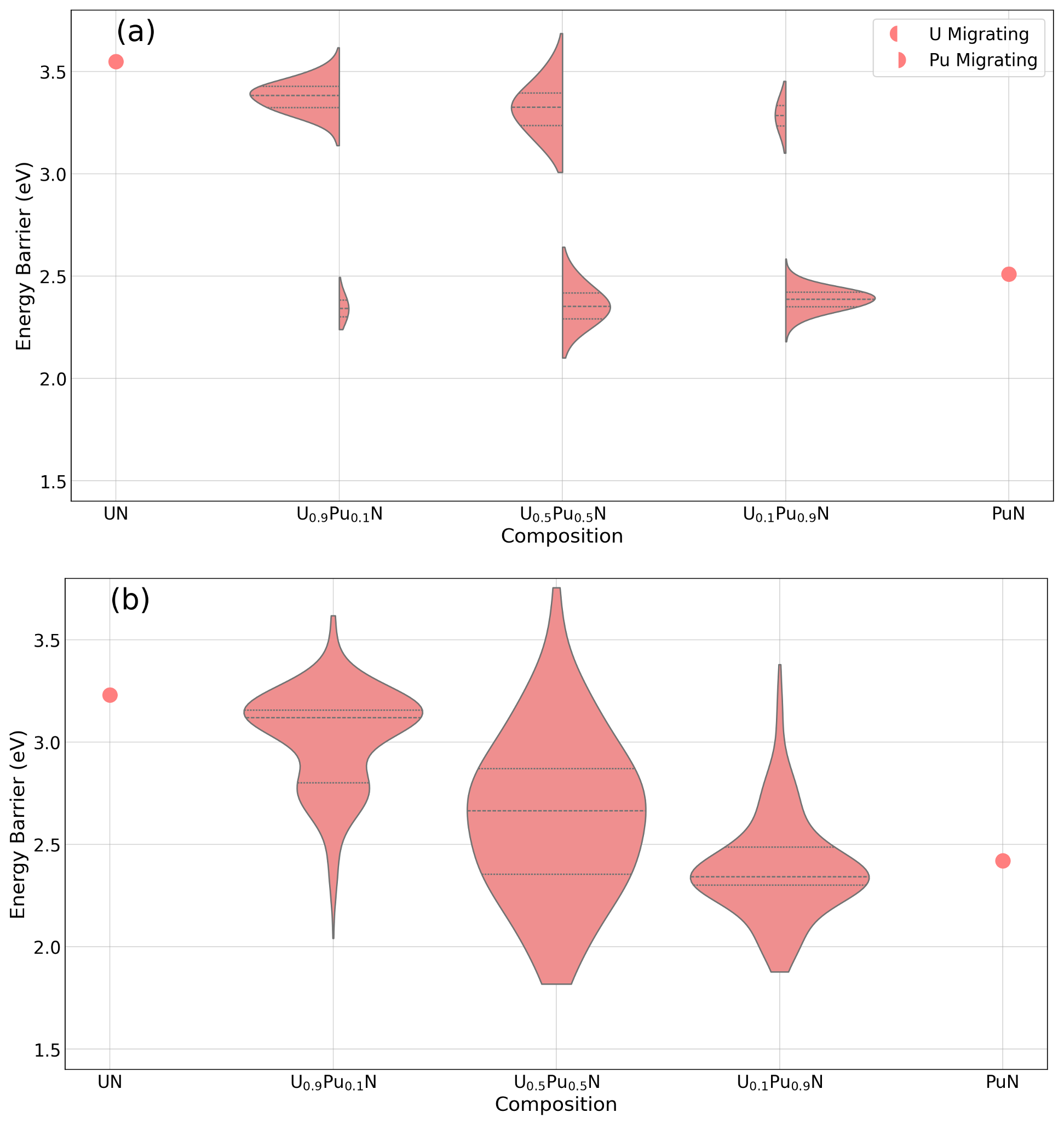}
    \caption{Energy barrier distributions for vacancy migration on the (a) actinide sublattice and (b) the nitride sublattice in structures with varying global concentrations of U/Pu derived from HopDec.}
    \label{fig:UN-defect-data}
\end{figure}

Figures \ref{fig:U-defect-data}–\ref{fig:UN-defect-data} present the distributions of migration barriers obtained from the HopDec workflow for each defect type at varying global U/Pu compositions. These distributions reflect the statistical sampling of distinct local chemical environments arising from the random redecorations.

For V$_\text{A}$ (Figure \ref{fig:U-defect-data}), the data are separated according to the identity of the migrating species, since the vacancy may be filled by either U or Pu depending on the local configuration. A clear bifurcation of the barrier distributions is observed between U-mediated and Pu-mediated hops. In both cases, the mean migration barrier closely matches that of the corresponding pure end member (UN or PuN), with only slight broadening due to local chemical variation. This indicates that the migration barrier on the actinide sublattice is governed predominantly by the identity of the migrating species rather than by the surrounding chemical environment. The local configurational disorder introduces a secondary spread in barrier height, but this effect is comparatively small. This behaviour is consistent with prior observations in chemically disordered alloys such as Cu–Ni, where vacancy migration barriers are strongly species-dependent and only weakly perturbed by local fluctuations \cite{Hatton2025}. In U$_x$Pu$_{1-x}$N, the principal compositional effect on actinide vacancy transport therefore arises from the relative frequency of U versus Pu mediated events, rather than from substantial environment-induced modulation of individual barriers.

In contrast, V$_\text{N}$ migration (Figure \ref{fig:N-defect-data}) involves only N as the migrating species, and therefore no bifurcation by species occurs. Instead, a continuous evolution of the barrier distribution with composition is observed. An overall reduction in the mean migration barrier is evident with increasing Pu content. At intermediate compositions (e.g. U$_{0.5}$Pu$_{0.5}$N), the distribution is relatively broad and unimodal, consistent with a wide sampling of mixed local environments. At low and high Pu concentrations, however, the violin plots exhibit more structured, multi-lobed features. These lobes reflect discrete classes of local actinide configurations surrounding the migrating N, corresponding to distinct combinations of U- and Pu-rich coordination shells. Interestingly, the mean barrier at U$_{0.1}$Pu$_{0.9}$N is lower than that of pure PuN. This suggests that a small concentration of uranium can introduce locally favourable environments that enable reduced-barrier migration pathways. In other words, dilute U in a Pu-rich matrix may create energetically advantageous configurations for N hopping, potentially giving rise to enhanced transport relative to the pure end member. This non-monotonic behaviour highlights the importance of explicitly sampling local chemical disorder rather than relying on the rule of mixtures between end-member energetics.

For the coupled V$_\text{A-N}$ defect (figure \ref{fig:UN-defect-data}), migration can proceed via either actinide or nitride motion while maintaining nearest-neighbour defect coordination. The barrier distributions for these two mechanisms are shown separately. In both sublattices, the qualitative shape of the distributions closely resembles those of the corresponding isolated vacancies (figures \ref{fig:U-defect-data} and \ref{fig:N-defect-data}). The presence of the coupled defect primarily induces a systematic shift in barrier height rather than a substantial restructuring of the distribution. This suggests that, to first order, the dominant chemical effects on migration remain sublattice-specific, with the coupled configuration modifying the energetic baseline but not fundamentally altering the sensitivity to local composition.

Overall, the HopDec dataset reveals two key features. First, actinide vacancy migration is strongly species-controlled, leading to composition-dependent kinetics driven by the statistical weighting of U and Pu hops. Second, nitride vacancy migration exhibits genuine environment-dependent modulation, including the emergence of favourable low-barrier configurations at certain intermediate compositions. These trends motivate the need for compositionally aware surrogate models capable of resolving both species effects and local chemical disorder.

\subsection{Surrogate Models}\label{sec:2}

\begin{table*}
    \centering
    \begin{tabular}{|c|c|c|c|c c|}
        \hline   
        \textbf{Fit} & \textbf{Parameter} & \textbf{V$_\text{A}$} & \textbf{V$_\text{N}$} & \textbf{V$_\text{A-N}$} & \textbf{V$_\text{A-N}$}  \\
         & & & & \textbf{A Mig.} & \textbf{N Mig.} \\ \hline
         $\Delta E$ &  $a_1$    &   -0.0156                         & 0.442         & 0.0260                & -0.465  \\
         $\Delta E$ &  $a_2$    &   -0.0479                         & -0.0113       & 0.0445                & 0.00452 \\
         $\Delta E$ &  $a_3$    &   0.000292                        & 0.00988       & -0.000516             & -0.00749 \\
         $\Delta E$ &  $a_4$    &   -0.0100                         & 0.00          & 0.0103                & 0.00 \\
         $E_b$      &  $b_1$    &   -0.00710, -0.0108               & 0.2179        & 0.0163, 0.0130        & -0.233 \\
         $E_b$      &  $b_2$    &   -0.0262, -0.0185                & 0.00          & 0.0239, 0.0197        & 0.00 \\
         $E_b$      &  $b_3$    &   2.02$\times 10^{-5}$, 0.000542  & 0.00          & -0.00203, -0.000958   & 0.00 \\
         $E_b$      &  $b_4$    &   -0.00614, -0.00562              & 0.00          & 0.00779, 0.00499      & 0.00 \\
         $E_b$      &  $c_1$    &   0.0763, 0.0723                  & -0.0982       & 0.0713, 0.0573        & -0.0541 \\
         $E_b$      &  $c_2$    &   -0.0366, -0.0119                & -0.140        & -0.0275, -0.0119      & -0.162 \\
         $E_b$      &  $c_3$    &   0.00635, 0.00131                & 0.183         & 0.00545, 0.00269      & 0.179 \\
         $E_b$      &  $c_4$    &   0.00454, -0.00853               & 0.218         & -0.00161, 0.00858     & 0.00 \\
         $E_b$      &  $\beta$  &   3.02, 2.25                      & 1.91          & 3.21, 2.37            & 2.28 \\
         \hline
         $\Delta E$ &  RMSE     &   0.0274                          & 0.0144        & 0.0317                & 0.0228 \\
         $E_b$      &  RMSE     &   0.0526, 0.0419                  & 0.0646        & 0.0703, 0.0530        & 0.0759 \\
         \hline
    \end{tabular}
    \caption{Fitted parameters for the surrogate models of $\Delta E$ and E$_b$ in equations (1) \& (2) with root mean-squared error (RMSE). Parameters for E$_b$ on the actinide (A) sublattice depend strongly on the migrating species so values are shown for both U and Pu respectively.}
    \label{tab-surrparam}
\end{table*}

Using the HopDec dataset described above, we fit the linear surrogate forms given in Eqs. (1) and (2) for each defect type, with fitted parameters shown in Table \ref{tab-surrparam}. For the V$_\text{A-N}$ defect, separate surrogate models are constructed for migration on the actinide and nitride sublattices. Furthermore, for barriers on the actinide sublattice, distinct fits are required for U- and Pu-mediated hops, reflecting the strong species dependence identified in Section III.A. A single parameterization is insufficient to accurately capture both cases simultaneously. Despite the chemical disorder inherent to U$_x$Pu$_{1-x}$N, the migration energetics are captured with relatively simple linear models based solely on local U counts within a small number of radial shells of the defect centre of mass. No higher-order interaction terms or non-linear descriptors are introduced. Nevertheless, the resulting root mean-squared errors (RMSEs) for both $\Delta E$ and E$_b$ are low across all defect types, indicating that the dominant contributions to migration energetics are well described by these coarse local chemical descriptors. The accuracy of a strictly linear model suggests that, to first order, migration energetics respond approximately additively to changes in local actinide identity. In other words, the chemical perturbation introduced by substituting U with Pu in successive coordination shells contributes nearly independently to the total barrier and energy change. Strongly non-linear or cooperative effects appear to be subdominant for the system considered here.

The functional form for $\Delta E$ (Eq. 1) depends only on changes in U coordination between the initial and final minima which implies that the thermodynamic driving force for a hop is controlled by how the local U/Pu balance changes during migration. The fitted coefficients $a_k$ therefore quantify the energetic preference of the defect for U-rich versus Pu-rich environments within each coordination shell. A positive $a_k$ indicates that increasing U content in shell $k$ (Table I) stabilizes the final state relative to the initial state, while a negative value indicates the opposite.

The barrier expression (Eq. 2) contains two physically distinct contributions. The $b_k$ terms belong to the same minimum-to-minimum coordination changes as $\Delta E$ and therefore represent how asymmetry between initial and final environments modifies the activation barrier. In contrast, the $c_k$ coefficients depend explicitly on the U content at the saddle point. These terms can be interpreted as describing how the transition state energy is stabilized or destabilized by the local chemical environment independently of the initial and final minima. The constant term $\beta$ corresponds to a baseline migration barrier.

For migration on the actinide sublattice, the fitted parameters differ markedly between U and Pu mediated hops. This reflects the bifurcation observed in Figure 1 and confirms that the identity of the migrating atom dominates the barrier height. In practice, the surrogate model separates this strong species effect (encoded primarily in $\beta$) from the weaker environmental modulation captured by the coordination terms. The comparatively small RMSE values for each species specific fit indicate that once species identity is accounted for, residual environmental effects are modest and approximately linear. In contrast, for V$_\text{N}$ migration, only a single species migrates and the environment-dependent coefficients play a comparatively larger role. For the coupled V$_\text{A-N}$ defect, separate fits for actinide and nitride motion demonstrate that the chemical sensitivity largely remains sublattice-specific, with the presence of the paired vacancy introducing a systematic shift rather than fundamentally altering the functional dependence on local coordination.

A key feature of the surrogate approach is its computational efficiency for evaluation. The models use a maximum of four coordination shells and simple integer counts of U atoms as descriptors. No explicit geometric distortions, higher-order chemical clusters, or machine-learned non-linear mappings are required. Given the relatively low RMSE values achieved, this suggests that the essential physics that the interatomic potential is capturing about defect migration in U$_x$Pu$_{1-x}$N is governed primarily by short-range chemical identity rather than long-range collective effects. This simplicity is advantageous for two reasons. First, it enables rapid evaluation of migration energetics during KMC simulations without expensive on-the-fly barrier calculations. Second, it provides a physically interpretable framework in which the effect of composition can be traced directly to changes in local coordination statistics. Overall, the low fitting errors combined with the compact functional form indicate that migration energetics in compositionally disordered actinide nitrides admit an unexpectedly simple reduced-order representation. 

\subsection{Kinetic Monte Carlo (KMC) Simulations}\label{sec:3}

The fitted surrogate models are implemented in a lattice-based kinetic Monte Carlo (laKMC) simulator to compute long-time defect diffusivities for each defect type as a function of temperature and composition across U$_x$Pu$_{1-x}$N. For each composition, simulations are performed over multiple random U/Pu decorations and multiple independent trajectories per decoration, allowing composition-dependent diffusivities to be estimated with statistical averaging. Diffusivities are extracted from the long-time slope of the mean-squared displacement using Eq. (4). The resulting trends are shown in Figures 4-6.

\begin{figure*}
    \centering
    \includegraphics[width=\linewidth]{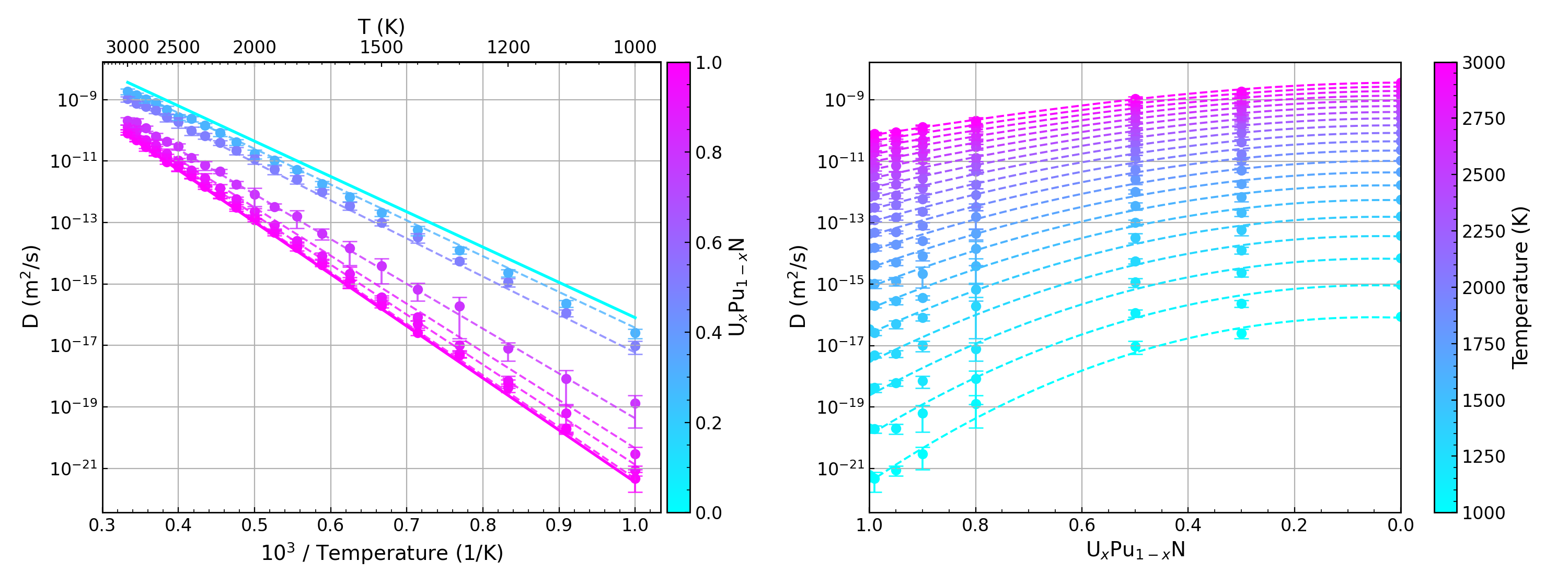}
    \caption{Diffusivity of a vacancy on the actinide sublattice as a function of (a) Temperature and (b) Concentration.}
    \label{fig:KMC_UVac}
\end{figure*}

Figure 4(a) for the V$_\text{A}$ defect shows Arrhenius behaviour over the simulated temperature range, with a clear composition-dependent vertical offset and (in places) modest changes in effective activation slope. This is consistent with actinide vacancy migration being strongly species-controlled: U-mediated hops remain comparatively slow, while Pu-mediated hops are substantially faster. The composition dependence in Figure 4(b) is highly non-linear. At low Pu fractions (i.e., U-rich compositions), introducing a small amount of Pu produces a disproportionate increase in diffusivity. Physically, even dilute Pu introduces locally fast vacancy hops that can markedly increase the overall mobility when they occur frequently enough to influence the long-time random walk. As Pu concentration increases further, diffusivity continues to rise but begins to approach a plateau, consistent with the emergence of a sufficiently connected network of Pu-enabled low-barrier events. Beyond this point, adding more Pu yields diminishing returns because the rate-limiting steps increasingly cease to be \lq finding' a fast hop and instead become controlled by other factors such as the remaining slower segments of the trajectory. This behaviour is most pronounced at lower temperatures (Figure 4(b), lower curves), where the exponential sensitivity to barrier differences amplifies the contrast between U- and Pu-mediated events. At high temperatures, the disparity in hop rates is reduced, and composition has a weaker effect on the overall diffusivity because a broader range of barriers becomes dynamically accessible. Overall, actinide vacancy transport in U$_x$Pu$_{1-x}$N is controlled by the statistical availability and connectivity of fast Pu-mediated hops rather than by a smooth interpolation between UN and PuN end members.

\begin{figure*}
    \centering
    \includegraphics[width=\linewidth]{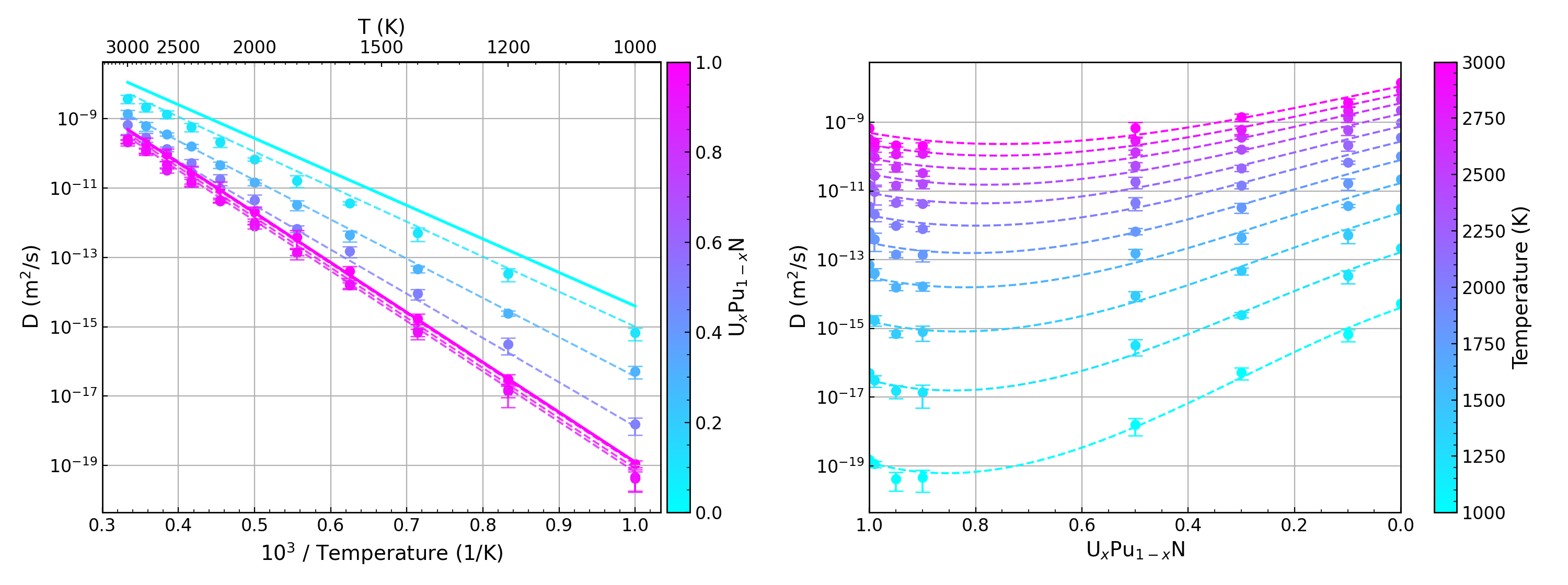}
    \caption{Diffusivity of a Nitride vacancy as a function of (a) Temperature and (b) Concentration.}
    \label{fig:KMC_NVac}
\end{figure*}

V$_\text{N}$ diffusion displays a qualitatively different dependence on composition and temperature. In Figure 5(a), the Arrhenius trends remain broadly linear in 1/T, but the separation between curves with different compositions is smaller and less monotonic than for the actinide vacancy case. This is consistent with the HopDec barrier distributions (Figure 2), where V$_\text{N}$ barriers vary continuously with composition rather than splitting cleanly by migrating species. In Figure 5(b), diffusivity shows a non-trivial response to Pu addition. In the U-rich regime, introducing a small amount of Pu can reduce diffusivity, indicating that Pu creates local environments that slow V$_\text{N}$ migration, effectively acting as kinetic traps. In KMC terms, these environments increase the residence time of the defect in certain states and reduce the net transport rate even if some lower-barrier events exist elsewhere. This is a typical finding in chemically complex systems and is typically referred to as sluggish diffusion \cite{Osetsky2018,Tsai2013,Dkabrowa2016}. At higher Pu contents, diffusivity increases again and eventually approaches the Pu-rich limit, consistent with the overall reduction in the mean V$_\text{N}$ barrier with increasing Pu content, while still reflecting the influence of barrier heterogeneity and the presence/absence of particularly favourable local configurations. Notably, the KMC results are not necessarily consistent with the observation from Figure 2, that certain mixed compositions can exhibit low-barrier subpopulations than PuN. However, whether these favourable events dominate macroscopic diffusion depends on their connectivity: isolated fast hops do not necessarily translate to high diffusivity if the defect must repeatedly traverse higher-barrier regions to make long-range progress. Thus, the composition dependence in Figure 5(b) reflects a competition between the emergence of low-barrier local configurations and the extent to which Pu introduces trapping environments that increase residence times. This provides a cautionary tale of conflating mean energy barrier directly with diffusivity.

\begin{figure*}
    \centering
    \includegraphics[width=\linewidth]{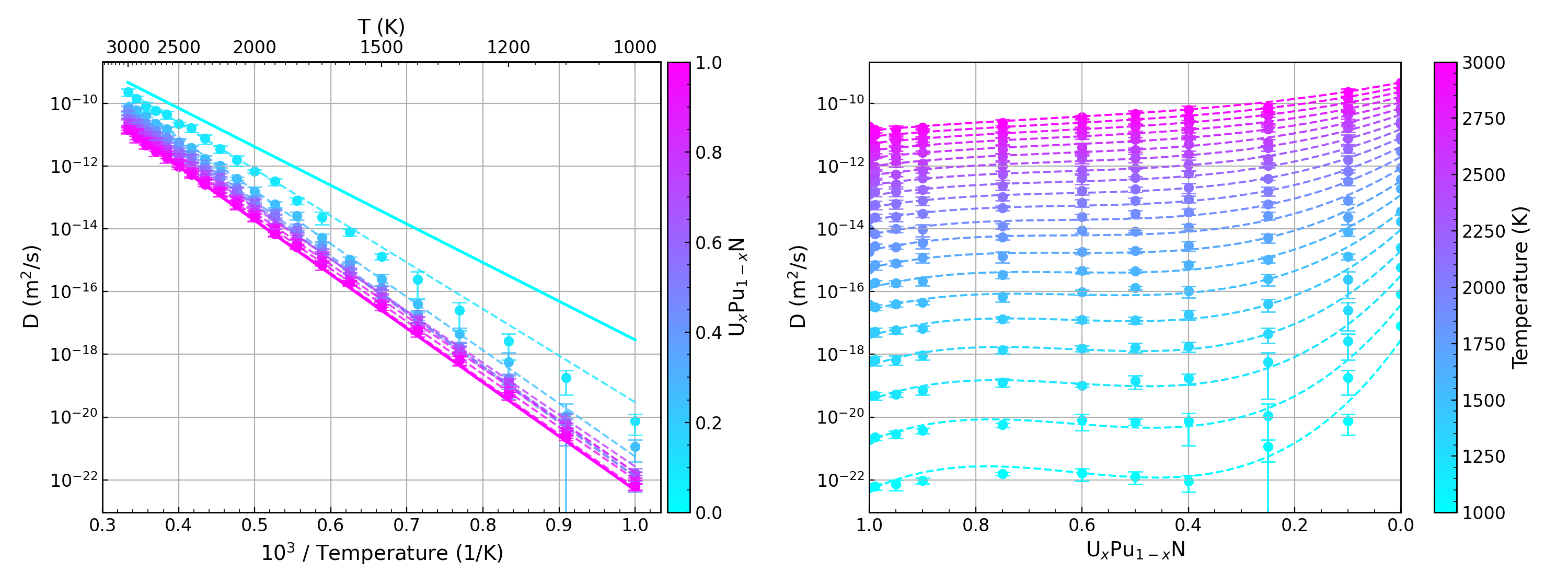}
    \caption{Diffusivity of an actinide-nitride coupled vacancy as a function of (a) Temperature and (b) Concentration.}
    \label{fig:KMC_UNVac}
\end{figure*}

For the coupled V$_\text{A-N}$ defect, diffusivities are shown in Figure 6. In these simulations, the divacancy is constrained to remain as a NN pair, enabling clean extraction of a single effective diffusivity for the bound complex. This approach isolates the migration kinetics of the pair itself, while association/dissociation processes can be treated separately in a higher-level model (e.g., cluster dynamics) \cite{Matthews2019}. Figure 6(a) again shows near-Arrhenius behaviour. Compared to isolated vacancies, the divacancy diffusivity is generally reduced, as expected for a correlated complex whose motion requires coordinated sequences of vacancy hops while maintaining the bound geometry. The composition dependence in Figure 6(b) is non-linear and reflects the fact that migration can proceed via either actinide or nitride motion within the pair, with each sublattice contribution exhibiting its own chemical sensitivity. In mixed compositions, the effective mobility therefore depends on which constituent hop types dominate the displacement of the bound pair and on how frequently the pair encounters environments that either facilitate or impede those constituent hops. As a result, its composition dependence can differ from either isolated V$_\text{A}$ or isolated V$_\text{N}$, even when the underlying barrier distributions resemble shifted versions of the isolated-vacancy cases (as suggested in Figure 3). In particular, small changes in composition can modify which hop mechanism is rate-controlling for the pair and can either enhance or suppress mobility depending on whether the controlling steps become faster and/or more connected.

Figures 4–6 demonstrate that diffusivity in U$_x$Pu$_{1-x}$N is not well described by simple interpolation between UN and PuN end members. Instead:
\begin{enumerate}
    \item Actinide vacancy diffusion is dominated by species-controlled kinetics (U vs Pu hops), producing strong non-linearity and a rapid rise in mobility once sufficiently many fast events contribute to long-range motion.
    \item Nitride vacancy diffusion is governed by environment-sensitive disorder effects, with evidence for both trap-dominated regimes at low Pu additions, known as sluggish diffusion, and enhanced mobility at higher Pu where lower barriers become more prevalent and/or better connected, namely percolation.
    \item Divacancy diffusion reflects coupled sampling of both sublattices, showing behaviour that can be qualitatively distinct from isolated vacancy transport and emphasizing the importance of mechanism competition within the bound complex.
\end{enumerate}

Fits to these data derive the following equations for diffusivity as a function of temperature and concentration, x:

\begin{equation}
    D(T,x) = D_0(x)\exp\!\left(-\frac{E_a(x)}{k_\mathrm{B} T}\right).
    \label{eq:arrhenius}
\end{equation}

Defining $A(x) \equiv \ln D_0(x)$, the composition dependences are written as

\begin{align}
    A(x) &= (1-x)\,A_\mathrm{PuN} + x\,A_\mathrm{UN} \nonumber \\
           &\quad + x(1-x)\!\left[L_0^{A} + L_1^{A}(2x-1)\right],
    \label{eq:Ax} \\[6pt]
    E_a(x) &= (1-x)\,E_a^\mathrm{PuN} + x\,E_a^\mathrm{UN} \nonumber \\
           &\quad + x(1-x)\!\left[L_0^{E_a} + L_1^{E_a}(2x-1)\right].
    \label{eq:Eax}
\end{align}

where $k_B$ is Boltzmann's constant, T is temperature and x is from U$_x$Pu$_ {1-x}$N. Fitted parameters are given in Table \ref{tab:KMCfits} for each defect type and displayed as dashed lines in Figures 4-6. The composition dependences of $\ln D_0$ and $E_a$ in equations \ref{eq:Ax} and \ref{eq:Eax}, respectively, are represented using Redlich–Kister expansions, following the same mathematical form commonly used for excess properties in binary mixtures \cite{Moggridge2012}. The first two terms provide a linear interpolation between the UN and PuN end members, a transport-coefficient analogue of Vegard’s law. However, the diffusivities obtained from the KMC simulations (Figures 4–6) deviate significantly from this ideal behaviour. The $x(1-x)$ terms represent the non-linear composition effects, vanishing at the end-member compositions. In a random solid solution, the statistical weight of mixed U-Pu neighbourhoods is proportional to $x(1-x)$. Since the migration kinetics are governed by the statistics of these mixed local environments, departures from linear interpolation are expected to follow a similar trend. The coefficient $L_0$ captures the leading-order influence of chemical disorder on transport, while the $L_1(2x-1)$ term is the asymmetric correction, allowing the composition dependence to differ between the U-rich and Pu-rich regimes. Although originally developed for thermodynamic excess functions, the Redlich–Kister formalism is employed here as a physically motivated representation of the non-linear transport behaviour arising from environment-dependent defect migration in U$_x$Pu$_{1-x}$N.

\begin{table}[h]
\centering
\begin{tabular}{|c|c|c|c|}
\hline
Parameter & N vacancy & U vacancy & UN divacancy \\
\hline
$A_\mathrm{PuN}$  & $-10.8742$ & $-10.6186$ & $-12.0541$ \\
$A_\mathrm{UN}$   & $-10.3811$ & $-10.4060$ & $-11.8341$ \\
$L_0^{A}$         & $-4.6352$  & $-2.7075$  & $6.1268$   \\
$L_1^{A}$         & $0.7385$   & $-1.4794$  & $-9.7858$  \\
\midrule
\hline
$E_a^\mathrm{PuN}$ & $1.9204$ & $2.2787$ & $2.4413$ \\
$E_a^\mathrm{UN}$  & $2.8558$ & $3.3592$ & $3.4112$ \\
$L_0^{E_a}$        & $0.5715$ & $-1.4575$ & $2.0965$ \\
$L_1^{E_a}$        & $0.8784$ & $-0.1697$ & $-3.0415$ \\
\hline
\end{tabular}
\caption{Fitted parameters for the three defect
         types in $\mathrm{U}_x\mathrm{Pu}_{1-x}\mathrm{N}$.
         Parameters $A$ are dimensionless (natural logarithm of $D_0$
         in m$^2$\,s$^{-1}$); activation energy parameters are in eV.}
\label{tab:KMCfits}
\end{table}

\subsection{Trapping Analyses}\label{sec:4}

The diffusivities presented in the previous section are emergent properties of defect motion through a chemically disordered energy landscape. While the KMC simulations quantify the resulting overall transport behaviour, they do not directly explain the atomic-scale origin of the observed composition dependence. In particular, variations in diffusivity may come from differences in defect trapping, residence times, or the connectivity of favourable migration pathways.

The surrogate models for $E_b$ and $\Delta E$ can also be used to understand these effects. By evaluating migration barriers and energy differences across a wide range of local environments, they enable direct analysis of defect escape and trapping behaviour that would be computationally prohibitive using explicit atomistic calculations alone. The aim of this section is therefore to connect the long-range diffusivities obtained from KMC to the underlying kinetic and thermodynamic landscape sampled by the defects.


\begin{figure*}
    \centering
    \includegraphics[width=\linewidth]{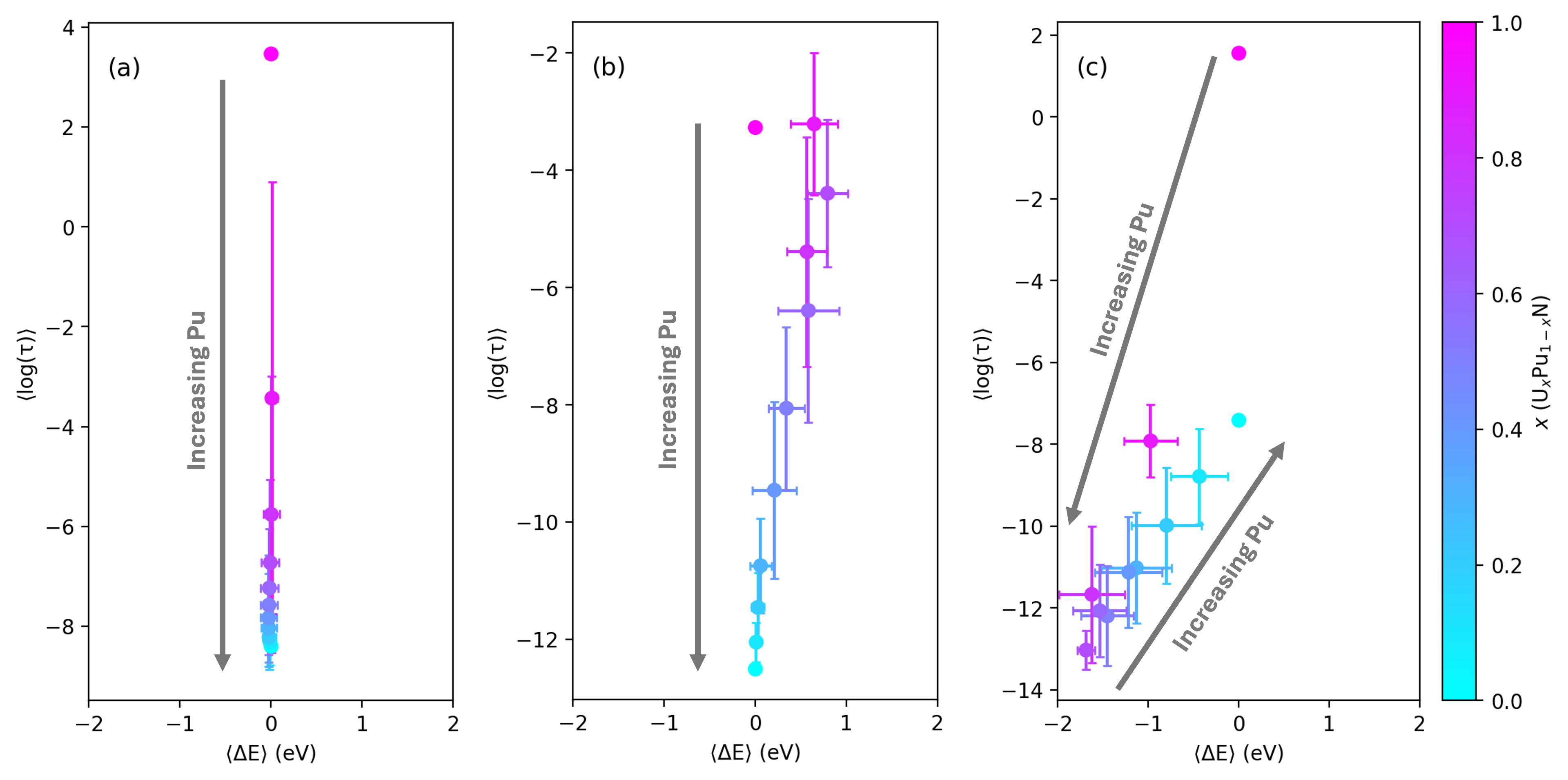}
    \caption{Two-dimensional kinetic–thermodynamic space showing Boltzmann-averaged residence time and $\Delta E$ per decoration for (a) Actinide vacancy, (b) Nitride vacancy, and (c) Actinide-Nitride coupled vacancy complex for U$_x$Pu$_{1-x}$N with equally spaced concentrations. Increasing Pu content indicated by arrow. The absolute magnitude of $\langle \tau \rangle_c$ should not be directly compared across different defect species, as each defect samples a distinct escape network and kinetic landscape.}
    \label{fig:trapping}
\end{figure*}

For each randomly decorated configuration $d$ with global composition $c$, all nearest-neighbour migration events $i = 1,\dots,M_d$ that move the defect out of state $s$ are identified. For each event $i$, we compute the activation barrier $E^{\ddagger}_{d,i}$ and $\Delta E_{d,i}$ using the previously discussed surrogate models. The total escape rate from state $s$ in $d$ is then,
\begin{equation}
    k_d(T) = \sum_{i=1}^{M_d} k_{d,i}(T)
\end{equation}
and the corresponding mean residence time is
\begin{equation}
    \tau_d(T) = \frac{1}{k_d(T)}
\end{equation}

To characterise the energetic directionality of escape, we define a rate-weighted mean energy change for decoration d,

\begin{equation}
\overline{\Delta E}_d(T)
=
\sum_{i=1}^{M_d} p_{d,i}(T)\,\Delta E_{d,i},
\quad
p_{d,i}(T) = \frac{k_{d,i}(T)}{k_d(T)},
\end{equation}

\noindent where $p_{d,i}(T)$ is the conditional probability that escape from decoration $d$ occurs via event $i$. This quantity represents the mean energetic change associated with the actual escape flux from state $s$ in environment $d$. To obtain composition-dependent observables representative of equilibrium sampling over local environments, we perform a Boltzmann average over decorations weighted by the absolute energy of the defect state:

\begin{equation}
    w_d(T) = \frac{\exp\!\left(-E_d/k_B T\right)}{\sum_{d'=1}^{N} \exp\!\left(-E_{d'}/ k_B T\right)}.
\end{equation}

Using these weights, we define the disorder-averaged mean residence time

\begin{equation}
    \langle \tau \rangle_c(T) = \sum_{d=1}^{N} w_d(T)\,\tau_d(T),
\end{equation}

\noindent and the disorder-averaged energetic bias of escape

\begin{equation}
\langle \Delta E \rangle_c(T) =\sum_{d=1}^{N} w_d(T)\,\overline{\Delta E}_d(T).
\end{equation}

Here, the subscript $c$ emphasizes that these quantities are evaluated at fixed composition. For each composition $c$, we compute the pair

\begin{equation}
    \Big( \langle \Delta E \rangle_c(T),\; \langle \tau \rangle_c(T) \Big),
\end{equation}

\noindent which defines a point in a two-dimensional kinetic–thermodynamic space. The quantity $\langle \tau \rangle_c$ characterizes the equilibrium-weighted average trapping time of the defect at the reference site, while $\langle \Delta E \rangle_c$ quantifies the equilibrium-weighted energetic bias of escape from that state. By repeating this procedure over compositions, we obtain a composition-dependent map of disorder-averaged defect escape dynamics. The quantities $\langle \tau \rangle_c$ and $\langle \Delta E \rangle_c$ define a two-dimensional representation of defect escape dynamics at fixed composition c. The former characterizes the equilibrium-weighted kinetic stability of the defect in the reference state, while the latter quantifies the energetic bias of the escape flux. Taken together, they provide a coarse-grained description of the local free-energy landscape experienced by the defect. It should be noted that the absolute magnitude of $\langle \tau \rangle_c$ should not be directly compared across different defect species, as each defect samples a distinct escape network and kinetic landscape.

Positive values of $\langle \Delta E \rangle_c$ indicate that escape events are, on average, energetically uphill, whereas negative values indicate downhill relaxation. For clarity, we discuss the physical meaning of the limiting regimes defined by the signs and magnitudes of $\langle \tau \rangle_c$ and $\langle \Delta E \rangle_c.$

\begin{equation}
    \langle \tau \rangle_c \ \text{large}, \qquad \langle \Delta E \rangle_c > 0.
\end{equation}

This regime corresponds to kinetically stable, thermodynamically favoured defect states. The defect resides for long times in low-energy environments, and escape requires an energetically uphill transition. Physically, this is characteristic of deep trapping behaviour. In this regime, transport is expected to be suppressed, and the dynamics are dominated by thermally activated release events.

\begin{equation}
    \langle \tau \rangle_c \ \text{large}, \qquad \langle \Delta E \rangle_c < 0
\end{equation}

Here, the defect remains kinetically long-lived in the reference state, but escape events are energetically downhill. This indicates the presence of rare, lower-energy neighbouring states that are separated from the reference configuration by relatively high barriers. Such behaviour is consistent with a funnel-like landscape: the defect is metastable in the reference state but, once activated, relaxes toward deeper basins.

\begin{equation}
    \langle \tau \rangle_c \ \text{small}, \qquad \langle \Delta E \rangle_c > 0
\end{equation}

In this regime, escape from the reference state occurs readily, but the typical transition leads to a higher-energy configuration. This suggests the existence of low-barrier pathways connecting to less stable states. Consequently, rapid forward transitions may be followed by frequent recrossing events, leading to high local transition activity without substantial net transport. This regime may therefore contribute to increased hop frequencies while only slightly enhancing long-range diffusivity typical of \lq flickering' events seen in many KMC simulations \cite{Adjanor2024}.

\begin{equation}
    \langle \tau \rangle_c \ \text{small} \qquad \langle \Delta E \rangle_c < 0
\end{equation}

This regime corresponds to kinetically unstable and thermodynamically unfavourable defect states. The defect escapes rapidly, and the transition is energetically downhill. Such configurations behave as transients in the energy landscape. This regime is expected to promote enhanced net mobility, as transitions are both kinetically mobile and energetically driven.

When $\langle \Delta E \rangle_c \approx 0$, the escape flux carries no net energetic bias, that is, the defect migrates without systematic preference for lower- or higher-energy states. Combined with a small $\langle \tau \rangle_c$, this is consistent with fast percolation through a relatively flat energy landscape. These four regimes provide broad heuristics for interpreting the composition-dependent kinetic–thermodynamic map. However, an important caveat is that because the analysis considers only the immediate escape events from the reference state, it cannot resolve extended multi-state basins. A composition that appears to lie near $\langle \Delta E \rangle_c = 0$ with moderate residence times could therefore reflect either genuine percolation or a super-basin structure, in which rapid local recrossing events rather indicate an overall trapping behaviour that would suppress net transport.

Applying this method to U$_x$Pu$_{1-x}$N, for each defect type we create this kinetic–thermodynamic space for 11 equally spaced $x$ values, the results are shown in Figure 7. Much of the data has large error bars and the majority of the time a consistent trend cannot be reliably identified. This is likely due to the complex heterogenous potential energy landscapes created by the compositional variations, however, a few interesting findings arise from analysis. 

For the V$_\text{A}$ defect, there is a simple relationship in which, with increasing Pu content, the residence time decreases and there is very little influence on the $\langle \Delta E \rangle_c$. This is consistent with Figure 4b showing a sharp increase in diffusivity correlating to a sharp decrease in residence time with small Pu concentrations. From this data, it can be claimed that adding Pu does not introduce traps, rather it produces fast pathways for percolation. 

In contrast, for the V$_\text{N}$ defect, we see that a small concentration of Pu leaves the residence time relatively unchanged but $\langle \Delta E \rangle_c$ increases. This implies that, small Pu concentrations create traps for V$_\text{N}$ by creating chemical environments which modulate relative energies rather than by increasing migration barriers. This is consistent with Figure 5b which indicates a reduction of net diffusivity of nitride vacancies at low Pu concentrations. Further, there is a steady decrease in $\langle \tau \rangle_c$ and $\langle \Delta E \rangle_c$ which converge to the PuN end member. This is once again consistent with Figure 5b which, beyond $x=0.9$, D quickly rises as residence times decrease and traps gives rise to faster migration pathways towards PuN.


Finally, for the V$_\text{A-N}$ defect, there is a complex kinetic-thermodynamic landscape. Small Pu concentrations induce a significant decrease in both $\langle \tau \rangle_c$ and $\langle \Delta E \rangle_c$. As discussed previously, this quadrant of the plot could be indicative of transient fast pathways in the energy landscape and indeed, there is an increase in D at these concentrations seen in Figure 6b. Beyond these concentrations there is relatively linear return to the end member of PuN. This does not exactly reflect the complexity seen in the diffusivity at these concentrations which is likely caused by a limitation of the approach taken here which only considers traps that are isolated and not extended as in basins. 

This approach gives significant insight into the trapping and long-range evolution of defects which is enabled by the kinetic and thermodynamic defect surrogates and would be unfeasible without them, since direct atomistic evaluation of escape rates and residence times across the full distribution of local environments at each composition would require a computationally intractable number of explicit barrier calculations.

\subsection{Discussion}\label{sec:5}

The results presented in Sections III.A–D demonstrate that defect transport in compositionally disordered U$_x$Pu$_{1-x}$N can be cast into a tractable, physically interpretable framework without sacrificing mechanistic accuracy. By combining migration data with linear surrogate models and a KMC simulator, we have coarse-grained a chemically complex diffusion problem into a compact kinetic description that retains the essential physics of migration, trapping, and escape in a disordered lattice.

A central finding is that compositional effects on defect transport manifest through two distinct mechanisms. The first is species-controlled kinetics, most evident for actinide vacancy diffusion, where the species of the migrating atom dominates the barrier height. The second is environment-dependant trapping and release, most notable in nitride vacancy migration, where barrier distributions are broad and local chemical configurations introduce both trapping sites and fast-diffusion channels. These mechanisms are not mutually exclusive; rather, they compete in determining long-range diffusivity. The KMC simulations show that global diffusion coefficients cannot be inferred from mean barrier values alone but emerge from the competition between barrier distributions, the connectivity of low-barrier pathways, and the statistics of residence times. The trapping analysis (Section IIID) provides a disorder-averaged projection of this through the $\langle\tau\rangle_c$–$\langle\Delta E\rangle_c$ representation, which effectively maps composition onto a reduced kinetic–thermodynamic landscape.

Several limitations are present, however, and potentially motivate future work. The actinide–nitride divacancy is constrained to remain as a nearest-neighbour pair during KMC, whereas in reality such complexes may undergo partial dissociation followed by re-association (a \lq break-away'/\lq catch-up' mechanism)\cite{Hatton2019}. Although dissociation is typically energetically uphill, thermal fluctuations at elevated temperatures could allow transient separation and correlated long-range displacement before re-binding, potentially enhancing effective mobility beyond that predicted for a strictly bound pair. Capturing this behaviour would require an object KMC treatment \cite{Mason2019} in which association and dissociation events are included with environment-dependent binding energies. In addition, only nearest-neighbour migration events are considered here; longer-range hops, correlated multi-atom rearrangements, or concerted exchange mechanisms may contribute at higher temperatures or under irradiation-induced nonequilibrium conditions \cite{Uberuaga2018,Vincent2006}. Additionally, while the low RMSE values of the surrogate fits suggest that NN effects dominate the sampled migration landscape, it is not clear how errors in barriers on the order of 0.05 eV impact the long-time trajectories of defects within this scheme since, at 1000 K this \lq small' error can cause a 2-times difference in the migration rate. Similarly, the Arrhenius prefactor $\nu$ is held constant, whereas in a chemically disordered systems the specific chemical species in the environment may strongly perturb this value. Extending the surrogate framework to parameterise $\nu$ alongside $E_b$ and $\Delta E$ as functions of local coordination would yield a more complete kinetic description.
Finally, in empirically based interatomic potentials, there is no explicit treatment of electronic effects, which may substantially impact the results presented here. Indeed, the presence of strongly localised 5f orbitals in Pu, compared to those in U, may introduce secondary effects to the energy landscape as a result of magnetism, the relative strength of this could change conclusions on preferential migration. Ideally, future work would introduce terms in the surrogate model derived from DFT calculations which explicitly parametrise the relative strength of the electronic effects. Alternatively, newly developed machine-learning frameworks may provide a path to include these effects in an interatomic potential.

The compact, reduced-order nature of the surrogate model makes it well-suited for integration into higher length and timescale methods. A natural next step is the development of an object KMC framework in which isolated vacancies, divacancies, and larger clusters are treated as interacting mobile objects with environment-dependent migration and binding rates. The surrogate functions derived here would provide rate evaluations within such a framework, while additional surrogates could describe cluster formation energies and emission barriers. Coupling this to a cluster dynamics simulator would enable the study of long-time defect population evolution under irradiation or thermal ageing. Because the surrogate formulation depends only on local chemical descriptors, it can naturally accommodate evolving compositions during burnup or transmutation, and the present work provides the necessary microscopic transport coefficients for single defects and small clusters across arbitrary material compositions.

Many technologically relevant materials, such as high-entropy alloys, doped semiconductors, solid electrolytes, and irradiated structural alloys also exhibit compositional disorder that breaks translational symmetry and generates distributions of defect energetics for which brute-force enumeration of local environments is intractable and/or computationally infeasible. In particular, in the case of nuclear fuels where local burnup depends on, for example, radial position in the fuel meat, these composition-aware diffusivities could be key to classifying the degradation mechanisms which occur at the complex high-burnup structures at the edge of the pellets.

A key result of this work is that migration energetics in such a chemically complex lattice can admit an unexpectedly simple reduced-order representation, here given by the Redlich-Kister expansion of composition in an Arrhenius form. Indeed, by treating chemical disorder as a parametrised statistical field rather than an explicitly enumerated configuration space, defect transport in evolving materials becomes computationally tractable without resorting to black-box machine learning models that lack physical interpretability. These findings highlight the importance of accounting for disorder-induced kinetic heterogeneity in predictive models of nuclear fuel behaviour and, more broadly, in the study of defect dynamics in chemically complex solids. \\

\section{Conclusion}

This work outlines a reduced-order framework for modelling defect diffusion in compositionally complex materials and applies it directly to representative defects in U$_x$Pu$_{1-x}$N mixed actinide nitrides. Directly calculated migration energetics are sampled across thousands of chemical environments for nearest neighbour migrations. Linear surrogate models for defect kinetics and thermodynamics are fitted to these data which, despite the chemical complexity of the system, resulted in models which reproduced the dataset with low RMSE and required only short-range species counts. This could indicate that the dominant contributions to defect migration energetics are simply governed by local chemical identity.

These surrogates are embedded in a KMC simulator which revealed that defect diffusivity in U$_x$Pu$_{1-x}$N exhibits strongly non-linear composition dependence that cannot be captured by simple Vegard's law type interpolation between UN and PuN end members. Actinide vacancy diffusion is primarily controlled by the identity of the migrating species, leading to disproportionate increases in mobility as fast Pu-mediated pathways percolate through the lattice. In contrast, nitride vacancy diffusion is governed by environment-dependent trapping and release processes arising from local chemical disorder. For the coupled actinide–nitride divacancy, diffusion reflects a competition between migration mechanisms on both sublattices, producing behaviour distinct from either isolated defect. Analysis of residence times and escape energetics further demonstrates how chemical disorder modifies local energy landscapes by introducing traps, metastable basins, and fast percolation pathways.

Outside of this application case, we demonstrate that defect migration in chemically complex materials can be coarse-grained into computationally tractable surrogate models while retaining physical interpretability. With the automated sampling of local environments through HopDec \cite{Hatton2025}, surrogate models of migration energetics, and long-range transport simulations, we have successfully demonstrated a practical approach which allows for atomic scale mechanisms to be up-scaled to transport equations tractable at the continuum/mesoscale. Such reduced-order descriptions are particularly valuable for multiscale simulations of nuclear fuel performance, in which evolving composition and defect populations must be represented efficiently over long timescales.

\begin{acknowledgments}
The author is grateful to Andrew Davies, Simon Middleburgh, Gareth Stephens, Aidan Cole-Baker, and Sam Armson for providing critical reviews of this manuscript.
The author acknowledges the helpful discussions with Blas Pedro Uberuaga and Danny Perez during the initial development of this workflow.
The author acknowledges the Hades computing cluster at Amentum for all computational work conducted in this study as well as Jim Skelton, Liam Marsden \& Andrew Dean for their help with this resource.
\end{acknowledgments}

\bibliography{bibliography}

\end{document}